\documentclass[%
sor,
jor,
amsmath,amssymb,
preprint,
]{revtex4-2}
\pdfoutput=1

\usepackage{graphicx}% Include figure files
\usepackage{dcolumn}% Align table columns on decimal point
\usepackage{bm}% bold math
\usepackage[mathlines]{lineno}% Enable numbering of text and display math
\begin{document}

\title[Dynamic Folding of Bistable Origami]{Dynamic Folding of Origami By Exploiting Asymmetric Multi-Stability}

\author{Sahand Sadeghi\footnote{Corresponding author: ssadegh@clemson.edu} and Suyi Li \\ Department of Mechanical Engineering, Clemson University, Clemson, SC USA}
% \altaffiliation[Corresponding author:{ssadegh@clemson.edu}
% \author{Suyi Li}
% \affiliation{Clemson University}

%\begin{linenumbers}

\begin{abstract}

In this study, we examine a rapid and reversible origami folding method by exploiting a combination of resonance excitation, asymmetric multi-stability, and active control.   The underlying idea is that, by harmonically exciting a multi-stable origami at its resonance frequencies, one can induce rapid folding between its different stable equilibria without the need for using responsive materials.  To this end, we use a bi-stable water-bomb base as an archetypal example.  Via numerical simulation based on a new dynamic model and experimental testing, we show that the inherent asymmetry of waterbomb bi-stability can enable dynamic folding with relatively low actuation requirements.  For example, if the water-bomb initially settles at its ``weak'' stable state, one can use a base excitation to induce the intra-well resonance.  As a result, the origami would fold and remain at the other ``strong'' stable state even if the excitation does not stop. The origami dynamics starting from the strong state, on the other hand, is more complicated.  The water-bomb origami is prone to show inter-well oscillation rather than a uni-directional switch due to a nonlinear relationship between the dynamic folding behavior, asymmetric potential energy barrier, the difference in resonance frequencies, and excitation amplitude.  Therefore, we develop an active feedback control strategy, which cuts off the base excitation input at the critical moment to achieve robust and uni-directional folding from the strong stable state to the weak one.  The results of this study can apply to many different kinds of origami and create a new approach for rapid and reversible (self-)folding, thus advancing the application of origami in shape morphing systems, adaptive structures, and reconfigurable robotics.

\end{abstract}


\maketitle

\section{Introduction}\label{Section 1}

Origami---the ancient art of paper folding---has received a surge of interests over the past decade from many research communities, such as mathematicians, material scientists, biotics researchers, and engineers (\cite{Li2019ArchitectedProperties}).  A key driving factor underneath such interests is the seemingly infinite possibilities of developing three-dimensional shapes from folding a simple flat sheet.  The \emph{kinematics} (or shape transformation) of origami is rich and offers many desirable characteristics for constructing deployable aerospace structures (\cite{Schenk2014ReviewRigidization}), kinetic architectures (\cite{Chen2015OrigamiPanels, Filipov2015}), self-folding robots (\cite{Felton2014AMachines}), and compact surgery devices (\cite{Randall2012Self-foldingApplications, Johnson2017FabricatingReview}). The \emph{mechanics} of origami offers a framework for architecting material systems (\cite{Li2019ArchitectedProperties}) with unique properties, like auxetics (\cite{Schenk2013GeometryMetamaterials}), tunable nonlinear stiffness (\cite{Fang2016a, Sadeghi2018TheMechanisms}), and desirable dynamic responses (\cite{Fang2017DynamicsStructure, Sadeghi2017HarnessingIsolation, Sadeghi2019FluidicIsolation}).  Moreover, the origami principle is geometric and scale-independent, so it applies to engineering systems of vastly different sizes, ranging from nanometer-scale graphene sheets (\cite{Cranford2009Meso-origami:Sheets}) all the way to meter-scale civil infrastructures (\cite{Ario2013DevelopmentSkill}).  

For most of these growing lists of origami applications, large amplitude and autonomous folding (or self-folding) are crucial for their functionality. However, achieving a (self-)folding efficiently and rapidly remains a significant challenge (\cite{Crivaro2016BistableActuation}).  To this end, we have seen extensive studies of using responsive materials to achieve folding via different external stimuli, such as heat (\cite{Tolley2014Self-foldingHeating}), magnetic field (\cite{Miyashita2015AnDegrades}), ambient humidity change (\cite{Okuzaki2013Humidity-sensitiveActuators}), and even light exposure (\cite{Ahmed2014InvestigatingStructures, Peraza-Hernandez2013Simulation-BasedSystem}).  In a few of these studies, bi-stability was also introduced as a mechanism to facilitate folding and maintain the folded shape without requiring a continuous supply of stimulation (\cite{Hanna2014, Bowen2015DevelopmentBase}). While promising, the use of responsive materials could incur complicated fabrication requirements, and their folding can be slow or non-reversible.

\begin{figure}[t!]
    \includegraphics[width=1.0\textwidth]{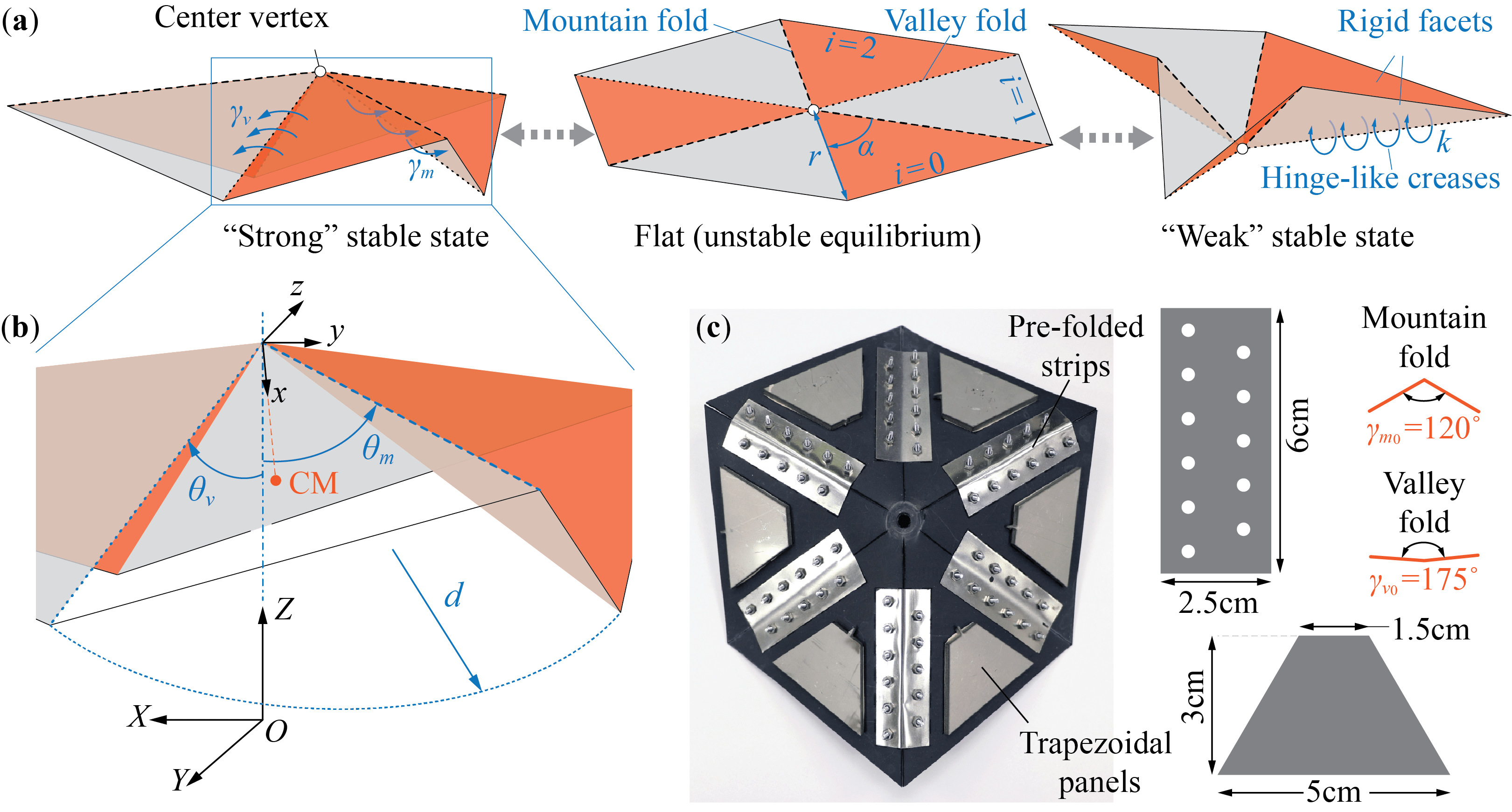}
    \centering
    \caption{The design, folding kinematics, and prototyping of the water-bomb base origami. (a) The external shape of the water bomb origami at its unfolded flat configuration and two stable states ($N=6$). We assume the triangular facet is rigid, and the fold lines behave like hinges with embedded torsional springs. (b) Variables that define the folding kinematics.  The inertial frame of reference ($XYZ$) is attached to the ground, and the body frame of reference ($xyz$) is attached to the facets. (c) Proof-of-concept prototype made out of a polypropylene sheet with perforations along the creases.  The geometry of the pre-folded shim stocks used to create stiffness along the creases is shown along with its folding angles for mountain and valley crease.  The geometry of the water-jet cut trapezoidal panels is also shown.}
    \label{fig:design}
\end{figure}

In this study, we examine a rapid and reversible origami folding method by exploiting the combination of harmonic excitation and embedded asymmetric bi-stability.  Bi-stable structures possess two distant stable equilibria (or ``stable states''), and this strong non-linearity can induce complex dynamic responses from external excitation, such as super-harmonics, intra/inter-well oscillations, and chaotic behaviors (\cite{Harne2017HarnessingSensing}).  These nonlinear dynamics have found applications in wave propagation control (\cite{Nadkarni2014DynamicsPropagation}), energy harvesting (\cite{Harne2013ASystems}), sensing (\cite{Harne2014ASensing}), and shape morphing (\cite{Arrieta2013DynamicComposites, Senba2010AFibers}). Here, shape morphing is particularly relevant to folding, so we used a proof-of-concept numerical simulation to demonstrate the feasibility of using harmonic excitation to induce folding in a bistable water-bomb base origami (\cite{Sadeghi2019AnalyzingOrigami, sadeghi2020dynamic}) (Figure \ref{fig:design}(a)).  The bi-stability of the water-bomb base is asymmetric (\cite{Hanna2014, Bowen2015DevelopmentBase}), so the resonance frequencies of its two stable configurations differ significantly.  It is possible that when the water-bomb origami is harmonically excited at the resonance frequency of its current stable state, it can rapidly fold to and remain at the other stable state.  Moreover, the required excitation magnitude by this dynamic folding method is smaller than static folding.

Building upon this proof-of-concept study, this study aims to obtain a comprehensive understanding of the harmonically-excited rapid folding via a combination of dynamic modeling, experimental validation, and controller design.  First, we formulate a new and nonlinear dynamic model of a generic water-bomb origami and conduct an in-depth examination into the relationships among the dynamic folding behaviors, potential energy landscape, resonance frequencies, and excitation amplitudes.  Since the bistability of water-bomb origami is asymmetric, we can designate its two stable states as ``strong'' or ``weak'' based on the magnitude of potential energy barriers between them.  Our simulation and experiment results show dynamic folding from the weak stable state to the strong one is relatively easy, but folding in the other direction is quite challenging to achieve.  That is, starting from the strong stable state, the water-bomb origami tends to exhibit inter-well oscillations under most excitation conditions, which is undesirable for rapid folding purposes. This challenge is further complicated by the fact that the nonlinear dynamics of origami are highly sensitive to design variations, fabrication errors, and excessive damping. Therefore, we then devise and experimentally validate a control strategy that ensures the robustness of dynamic folding by cutting off the excitation input at a critical configuration.  This control strategy is essential for practical implementations of this dynamic folding method in the future. 

It is worth highlighting that although this study uses the water-bomb origami as an example, the insights into the harmonically excited folding and the control strategy can apply to many other origami designs that exhibit asymmetric multi-stability, such as stacked Miura-ori (\cite{Fang2017AsymmetricStacked-origami}), Kresling (\cite{Cai2015}), and leaf-out pattern (\cite{Yasuda2016}). Moreover, harmonic excitation at the resonance frequency has a high actuation authority, so it can be an efficient method compared to other dynamic inputs, such as impulse (\cite{liu2018transformation}). Therefore, the results of this study can create a new approach for rapid and reversible (self-)folding, thus advancing the application of origami in shape morphing systems, adaptive structures, and reconfigurable robotics.

In what follows, Section 2 of this paper details the dynamic modeling of the water-bomb base origami, section 3 discusses its dynamic folding behavior under harmonic excitation, section 4 explains the active control strategy, and section 5 concludes this paper with a summary and discussion. 

\section{Dynamic Model of the Water-bomb Origami}\label{Section 2}

\begin{figure}
    \includegraphics[width=1.0\textwidth]{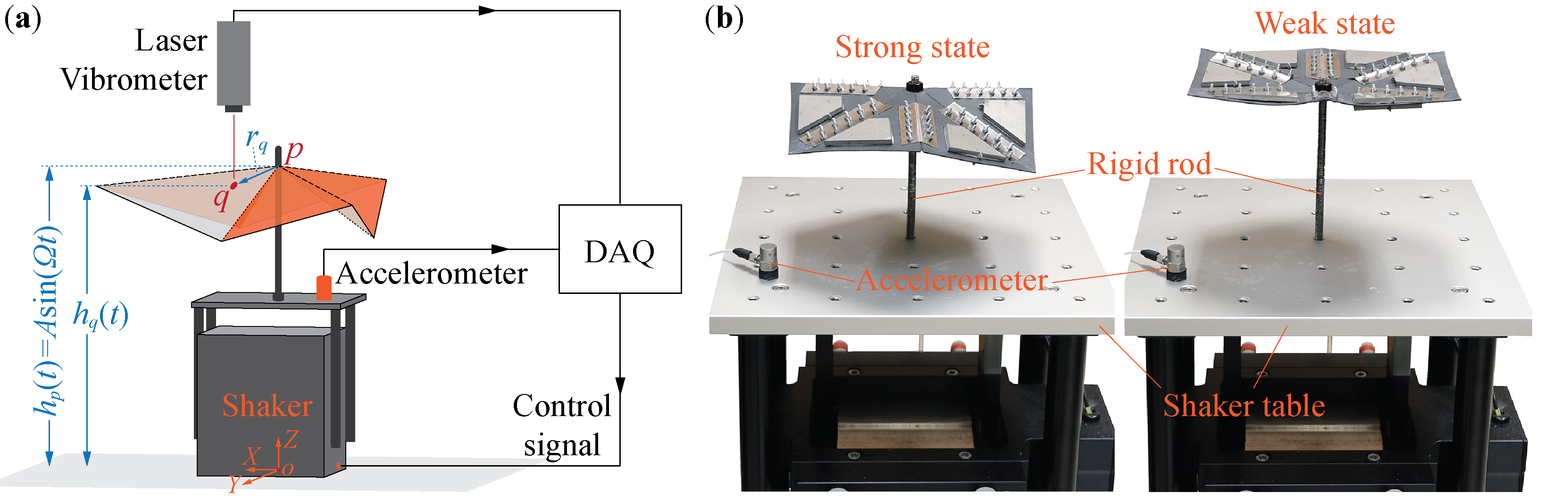}
    \caption{Dynamic folding test of the waterbomb origami. (a) A schematic drawing showing the overall experiment setup.  A rigid rod connects the central vertex to the shaker table.  The facets are free to rotate. The vertical oscillations of one of the facets are measured using a laser vibrometer ($r_q \approx 3$cm), which is then converted to folding angles in the DAQ system. The vibrations of the shaker are measured using a piezoelectric accelerometer. (b) The water-bomb origami prototype in its strong stable state (left) and weak stable state (right).}
    \label{fig:test}
\end{figure}

In this section, we derive the governing equation of motion of a generic water-bomb base origami.  Assuming the water-bomb is symmetric in its design and rigid-foldable (i.e., rigid facets and hinge-like creases), we can describe the kinematics of a water-bomb with $N$ triangular facets as a two degrees-of-freedom (DOF) mechanism.  These two degrees can be defined by the angle between the vertical $Z$-axis of the origami and its valley creases ($\theta_v$ in Figure \ref{fig:design}(b)), and the vertical position of the central vertex $h_p$, respectively.  We assume that this central vertex is rigidly connected to an external excitation, which is a vertical shaker table in this case (APS Dynamics 113, Figure \ref{fig:test}).  In this way, $h_p$ becomes the dynamic input variable, and $\theta_v$ is the only degree-of-freedom left. 

Using spherical trigonometry, we can derive the angle between the vertical $Z$-axis and the mountain creases of the structure as a function of $\theta_v$ in that:
\begin{equation} 
\theta_m=\cos^{-1} \left(\frac{\cos\alpha}{\cos (d/2)} \right)+\cos^{-1} \left(\frac{\cos\theta_v}{\cos (d/2)} \right),
\label{eq1}
\end{equation}
where $\alpha=2\pi/N$, and $d$ is the radius of a circular arc defined by the central vertex and two adjacent vertices on the valley creases (Figure \ref{fig:design}(b)):
\begin{equation} 
d=\cos^{-1} \left (\cos^2\theta_v+\sin^2\theta_v\cos\beta \right ),
\label{eq2}
\end{equation}
where $\beta=2\alpha$. Again, using spherical trigonometry, one can show that:
\begin{equation} 
\gamma_{m}=\pi-\cos^{-1} \left (1+\frac{\cos^2\theta_v+\sin^2\theta_v\cos\beta-1}{\sin^2\alpha} \right ),
\label{eq3}
\end{equation}
\begin{equation} 
\gamma_{v} = 
\left\{
\begin{aligned} 
& -\pi+2\cos^{-1}\left(\cot\alpha\tan\frac{d}{2}\right)+2\cos^{-1}\left(\cot\theta_v\tan\frac{d}{2}\right) & \text{if $\theta_v\leq\frac{\pi}{2}$} \\ 
&\pi-2\cos^{-1} \left (\frac{(\cos d-1)\cot\theta_v}{\sin d} \right )+2\cos^{-1}\left(\cot\alpha\tan\frac{d}{2}\right) & \text{if $\theta_v>\frac{\pi}{2}$} \label{eq4}
\end{aligned}
\right.
\end{equation}
where $\gamma_{m}$ and $\gamma_{v}$ are the angles between the facets connected by the mountain and valley creases, respectively (Figure \ref{fig:design}).

The position and orientation of each triangular facet can be described by the position of the central vertex $h_p$ in the $XYZ$ (inertial) frame of reference attached to the ground and the orientation of the $xyz$ (body) frame of reference. The latter can be described by three independent Euler angles, which represent the consecutive rotations of the $XYZ$ (inertial) frame of reference needed to align it with the $xyz$ (body) frame of reference. The order of rotations is arbitrary. Here, we choose the zyx order (aka, the aircraft rotations) that consists of three steps: The first step is a rotation about the $Z$-axis by $\psi_i$, where $\psi_i= \frac{2\pi}{N}$. The second step is a rotation about the $y'$-axis (aka., $y$-axis of the rotated frame after the first step) by $\theta_i$, where $\theta_i=\frac{1}{2}(\pi-\theta_v-\theta_m)$. The third step is a rotation about the $x''$-axis (aka. $x$-axis of the rotated frame after the second step) by $\phi_i$, where
\begin{equation} 
  \phi_i =
  \left\{
    \begin{aligned}
      &\frac{\gamma_m}{2} & \text{if $i$ is even,}\\
      &2\pi-\frac{\gamma_m}{2} & \text{if $i$ is odd.}\\
    \end{aligned}   
  \right.
    \label{eq5}
\end{equation}

Here, the sub-index ``$i$'' $(i=0 \ldots N-1)$ labels the different triangular facets as defined in Figure \ref{fig:design}(a). Therefore, the total rotation matrix is a combination of these three steps in that $\bm{C}_i=\bm{\Phi}_i\bm{\Theta}_i\bm{\Psi}_i$, where:
\begin{align} 
\bm{\Phi}_i &= \begin{bmatrix}
1&0&0\\
0&\cos\phi_i&\sin\phi_i\\
0&-\sin\phi_i&\cos\phi_i
\label{eq6}
\end{bmatrix},\\
\bm{\Theta}_i &=  \begin{bmatrix}
\cos\theta_i&0&-\sin\theta_i\\
0&1&0\\
\sin\theta_i&0&\cos\theta_i
\label{eq7}
\end{bmatrix},\\ 
\bm{\Psi}_i&= \begin{bmatrix}
\cos\psi_i&\sin\psi_i&0\\
-\sin\psi_i&\cos\psi_i&0\\
0&0&1
\label{eq8}
\end{bmatrix}.
\end{align}

In addition, the angular velocity of the $xyz$ (body) frame of reference can be derived using:
\begin{equation}
    \bm{\omega}_i=\omega_{xi}\bm{\hat{i}}+\omega_{yi}\bm{\hat{j}}+\omega_{zi}\bm{\hat{k}},
    \label{eq9}
\end{equation}
where:
\begin{equation} 
    \left\{
    \begin{aligned}
      &\omega_{xi}=\dot{\phi}_i-\dot{\psi}_i\sin\theta_i,  \\
      &\omega_{yi}=\dot{\psi}_i\cos\theta_i\sin\phi_i+\dot{\theta}_i\cos\phi_i,  \\
      &\omega_{zi}=\dot{\psi}_i\cos\theta_i\cos\phi_i-\dot{\theta}_i\sin\phi_i. \\
    \end{aligned}
    \right.
    \label{eq10}
\end{equation}

\subsection{\label{Section 2_sub 1}Kinetic energy of the origami}
As the water-bomb base origami folds, its facets exhibit both translational and rotational motions with respect to the central vertex. One can show that the total kinetic energy of the origami structure originates from these two distinct motions based on the following equations:
\begin{equation}
    T_{tot}=\frac{1}{2} N m|\bm{v}_p|^2+\sum_{i=0}^{N-1} \left(\frac{1}{2}\bm{\omega}_i^\intercal\bm{I}\bm{\omega}_i+m\bm{v}_p\cdot\dot{\bm{\rho}}_{ci}\right),
    \label{eq11}
\end{equation}
where, $m$ is the mass of a triangular facet, $\bm{v}_p$ is the velocity of the central vertex. $\dot{\bm{\rho}}_c = \bm{\omega}\times\bm{\rho}_{ci}$, where $\bm{\rho}_{ci}$ is the vector of center of mass of each facet in $xyz$ (body) frame of reference. Note that $\bm{v}_p$ and $\dot{\bm{\rho}}_{ci}$ should be expressed in the same frame of reference which is possible using the total rotation matrix $\bm{C}$. Finally, the matrix $\bm{I}$ contains that moment of inertia of each facet around the central vertex in that
\begin{equation}
\bm{I}=
 \begin{bmatrix}
    I_{xx}&0&0\\
    0&I_{yy}&0\\
    0&0&I_{zz}
 \end{bmatrix},
 \label{eq12}
\end{equation}
where, $I_{xx}=\frac{3}{4} mr^2\sin^2\alpha$, $I_{yy}= \frac{6}{7}mr^2\cos^2(\frac{\alpha}{2})$, and $I_{zz}= \frac{1}{2} m r^2+ \frac{1}{12} m r^2 \cos^2(\frac{\alpha}{2})$.

\subsection{\label{Section 2_sub 2}Gravitational potential energy of the origami}
In order to derive the gravitational potential energy of the water-bomb origami, we need to calculate the location of the center of mass of each facet. One can show that the distance of each center of mass from the ground can be derived using the following relationship:
\begin{equation}
    Z_{cm}=h_p- \left (\frac{2}{3}r\cos\frac{\alpha}{2}\cos \left (\frac{\theta_m+\theta_v}{2} \right ) \right ).
\label{eq13}
\end{equation}
The corresponding gravitational potential energy is $V_G=NmZ_{cm}$.

\subsection{\label{Section 2_sub 3}Elastic potential energy of the origami}

Assuming that the triangular facets are rigid and the creases behave like hinges with embedded torsional springs, we can derive the elastic potential energy of the origami as
\begin{equation} 
V_{E}=\frac{N}{2}\left[k_{\gamma_m}\left(\gamma_{m}-\gamma_{m_{0}}\right)^2+k_{\gamma_v}\left(\gamma_{v}-\gamma_{v_{0}}\right)^2\right],
\label{eq15}
\end{equation} 
where $k_{\gamma_m}$ and $k_{\gamma_v}$ are the torsional stiffness coefficient of the mountain and valley creases, respectively. $\gamma_{m}$ and $\gamma_{v}$ are the dihedral folding angles of the mountain and valley creases (Eq. \ref{eq3} and \ref{eq4}). In addition, $\gamma_{m_{0}}$ and $\gamma_{v_{0}}$ are the corresponding stress-free dihedral angles.

\subsection{\label{Section 2_sub 4}Equation of motion}
The Lagrangian of the origami structure becomes $\mathcal{L}=T_{tot}-V_G-V_E$, and we can derive the governing nonlinear equations of motion using
\begin{equation}
    \frac{\text{d}}{\text{d}t}(\frac{\partial\mathcal{L}}{\partial\Dot{\theta}_v})-\frac{\partial\mathcal{L}}{\partial \theta_v}+F_{d}=0.
    \label{eq17}
\end{equation}

$F_{d}$ is the damping force generated along the origami creases, and we assume that it has a simple form of $c r \dot{\theta}_v$. Here, $c$ is the equivalent viscous damping coefficient, and $r$ is the length of each crease. 

\section{Dynamic folding of the bistable water-bomb origami}\label{Section 3}

The equation of motion (\ref{eq17}) can be solved numerically using MATLAB ODE45 solver to obtain the dynamic response to arbitrary base excitation inputs and initial conditions. We assume that the base excitation is harmonic in that $h_{p}=A \cos \Omega t$. By solving the equation of motion under small-amplitude excitations and performing a stroboscopic sampling over a range of excitation frequencies, we obtain the intra-well frequency response of the water-bomb origami near its two stable equilibria. In this way, we can identify the corresponding intra-well resonance frequencies.  

We analyze the accuracy of the origami dynamic model by comparing its predicted frequency responses near the two stable states and experimental measurements from a proof-of-concept prototype. This prototype has a hexagonal shape with a crease length of 10cm ($N=6$ and $r=10$cm, Figure \ref{fig:design}(c)). We cut a 0.76mm thick flame-retardant Polypropylene sheet and perforated the crease lines using an FCX2000 series GRAPHTEC flatbead cutting plotter to create the compliant base layer of the origami. A significant amount of the material along the creases is removed to reduce the damping as much a possible. The torsional stiffness along the creases are generated by attaching 0.127mm thick shim stocks, which are folded carefully to give the initial stress-free crease hihedral angles $\gamma_{m_{0}}=120^\circ$ and $\gamma_{v_{0}}=175^\circ$.  Then, we attach twelve water jet-cut stainless steel trapezoids (24g each) to the triangular facets to offer sufficient inertia. Moreover, these trapezoids provide the desired rigidity to the facets according to the rigid-folding assumptions.   

\subsection{Intra-well frequency response analyses and parameter estimation}\label{Section 3_sub 1}

Although the geometric design and mass of the origami are known, we need to estimate the magnitudes of torsional stiffness ($k_{\gamma_m}$ and $k_{\gamma_v}$) and damping coefficient ($c$) of the creases. To this end, we perform intra-well frequency sweeps near both of its stable states with a small excitation amplitude to obtain the frequency response. Then we can estimate $k_{m}$ and $k_{v}$, which are assumed equal in this case, and $c$ by fitting the model predicted frequency responses to experimental results using the least square method.  In what follows, we show that these stiffness coefficients are crucial for determining the intra-well resonance frequency, and the damping coefficient directly affects the excitation amplitude for dynamic folding. 

\begin{figure}[t!]
    \includegraphics[]{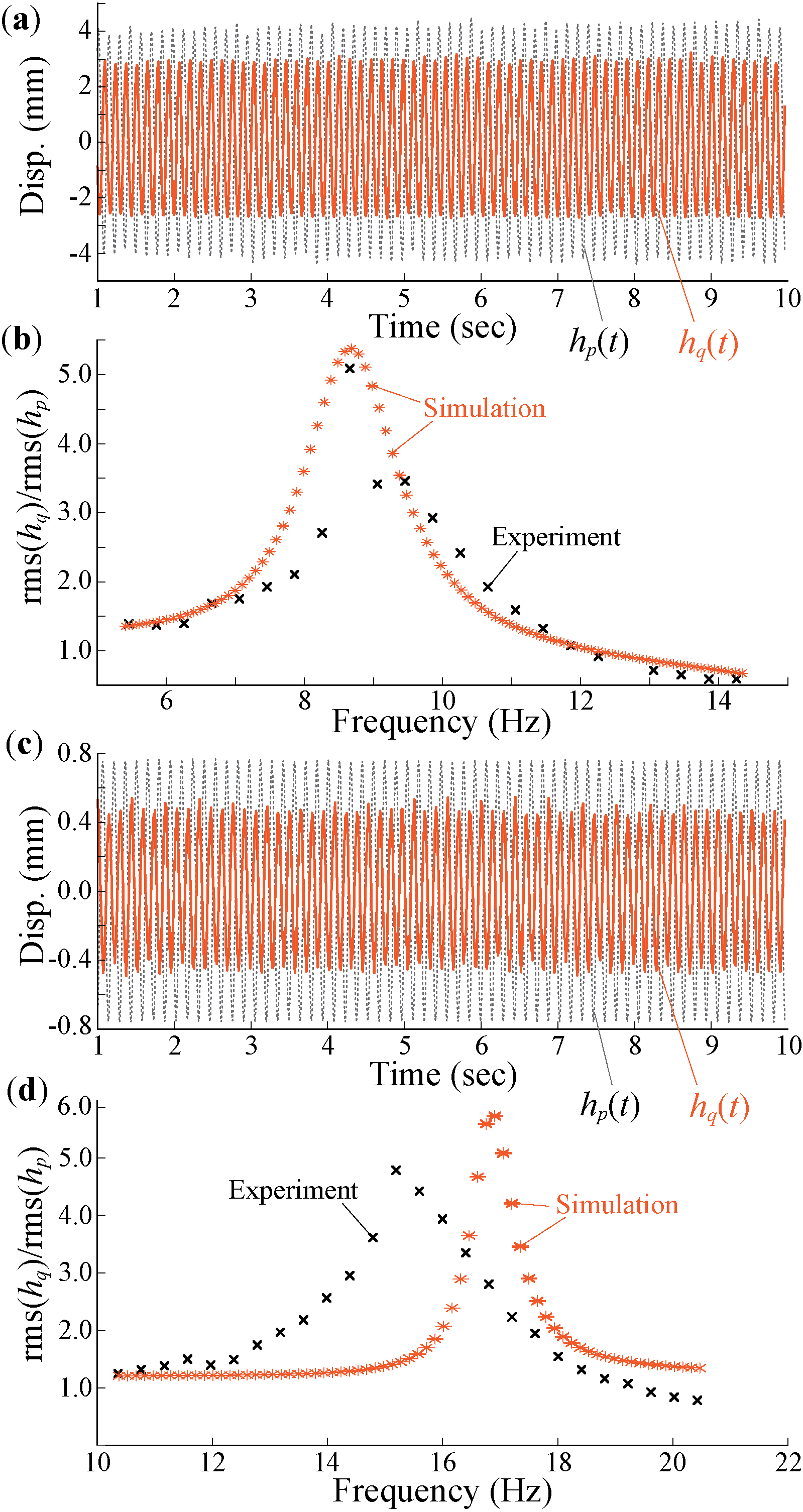}
    \caption{Frequency response near the two stable states of water-bomb origami. (a) A typical time response from small-amplitude intra-well oscillations around the stress-free stable state. Here, the vertical displacements of point $q$ ($h_{q}$) are represented in orange and the corresponding shaker excitations represented in gray.  (b) Stroboscopic sampling results for intra-well oscillations around the stress-free stable state. (c) Typical time response of similar small-amplitude intra-well oscillations around the other stable state. (d) The corresponding stroboscopic sampling results.}
    \label{fig:resonance}
\end{figure}

Figure \ref{fig:resonance}(a) shows the experimentally measured frequency response of the water-bomb origami near its stress-free stable state and the closest numerical prediction based on the least square method. Here, the frequency response is defined as $\frac{\text{rms}(h_q(t))}{\text{rms}(h_p(t))}$ after the transient response has damped out, where $h_{q}$ is the vertical displacement of a representative point on the median of a facet (Figure \ref{fig:test}(a)). We use an OFV-5000 Polytec laser vibrometer equipped with an OFV-503 laser head to capture the displacement response of this representative point and a 352C33 PCB accelerometer to measure the acceleration of the shaker, which is then converted to displacement. We find that $k_{\gamma_m}=k_{\gamma_v}=0.32$ $\frac{\text{N.m}}{\text{rad}}$ and $c=0.05$ $\frac{\text{kg}}{\text{rad.s}}$ give the best fitting at this stress-free stable state.  
Figure \ref{fig:resonance}(b) shows the experimentally measured frequency response near the other stable state and the corresponding numerical prediction by using the estimated crease stiffness and damping coefficient from the previous test. The comparison shows an approximately 15\% discrepancy between the estimated resonance frequency of this second stable state and the measured resonance frequency (15.8Hz based on the experiments and 17Hz based on simulation). This discrepancy probably originates from a combination of fabrication uncertainties and the simplifications made in the analytical model.  Moreover, the experimental results show higher damping than the prediction, which is reasonable due to the higher excitation frequencies at this stable state.  

\begin{figure}
    \includegraphics{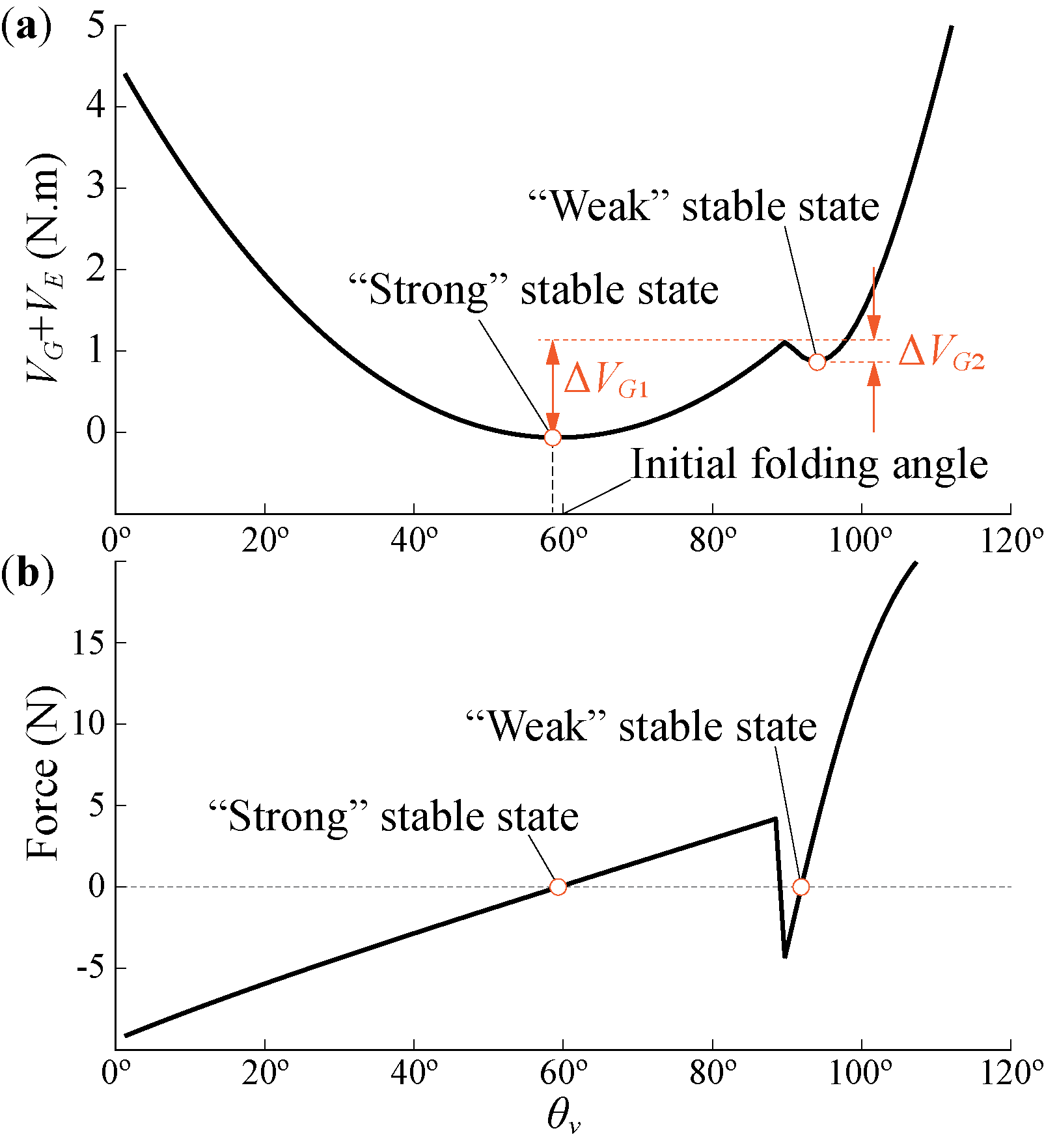}
    \caption{The asymmetric bi-stability of the water-bomb origami. (a) The asymmetric elastic potential energy landscape with two different energy barriers ($\Delta V_{G_1}$ and $\Delta V_{G_2}$). (b) The reaction force-displacement curve of the origami due to the elastic deformation along the creases.}
    \label{fig:FD}
\end{figure}

Overall, our model successfully captures the difference in the intra-well resonance frequencies near the two stable states of the water-bomb origami with a relatively small error. This difference in resonance frequencies comes from the inherently asymmetric potential energy landscape of the origami (Figure \ref{fig:FD}(a)), which creates an asymmetric force-displacement curve (Figure \ref{fig:FD}(b)) with different tangent stiffness near its two stable equilibria. For clarity, we refer to the initial, stress-free stable equilibrium with a deeper potential energy well as the ``strong'' state, and the other stable equilibrium with a shallower energy well as the ``weak'' state. The differences in the energy barriers for switching between these two stable states are evident.  That is, the origami must overcome a large barrier to switch from the strong stable state to the weak one, but only needs to overcome a small barrier for the opposite switch ($\Delta V_{G_1} > \Delta V_{G_2}$ in Figure \ref{fig:FD}).  In the following subsections, we show that the differences in resonance frequencies, energy barriers, and the base excitation amplitude all play crucial roles in the harmonically excited folding of water-bomb origami. 

\subsection{Dynamic folding from the weak stable state}\label{Section 3_sub 2}

If the water-bomb origami initially settles at the weak stable state, we can induce an intra-well resonance by exciting it with the corresponding resonance frequency.  In this way, the origami can exhibit a large reciprocal folding motion with a small energy input.  If the excited origami can overcome the energy barrier $\Delta V_{G_2}$, it can rapidly switch to the other, strong stable state.   Moreover, once this switch is complete, the water-bomb will remain in the strong state because 1) the energy barrier of the opposite switch is significantly higher, and 2) the resonance frequency of the strong state is significantly different from the original input frequency (that is, the intra-well resonance stops after the switch).

\begin{figure}
    \includegraphics[width=1.0\textwidth]{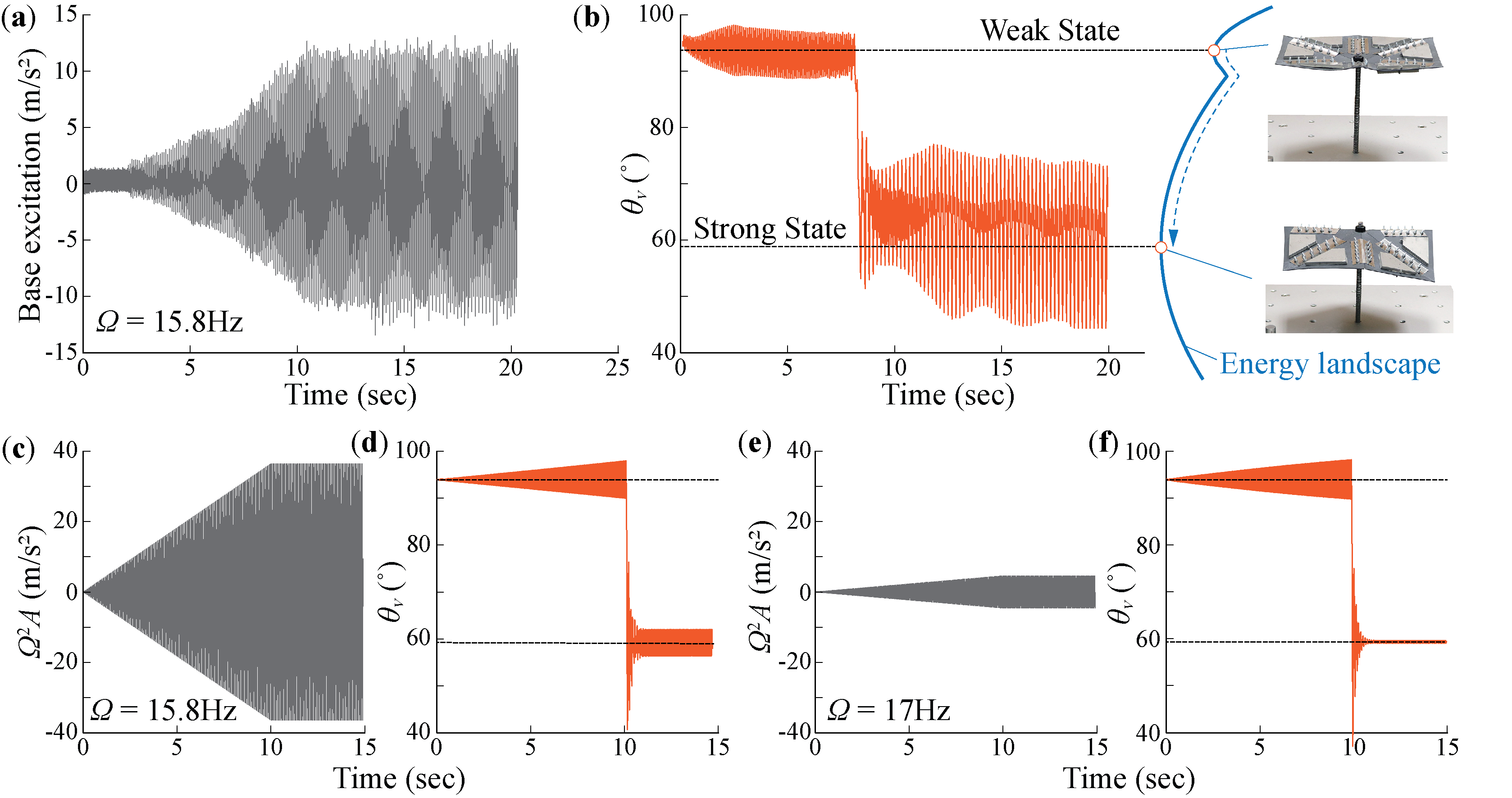}
    \caption{Harmonically excited folding from the weak stable state to the strong state. (a) Acceleration of the shaker's base (or central vertex) based on the piezoelectric accelerometer readings.  The frequency of excitations here is the experimentally measured resonance frequency from the actual prototype (15.8Hz).  The energy landscape is shown on the right for clarity. (b) Time response of $\theta_{v}$ calculated from that laser vibrometer data. (c) The base acceleration from the numerical simulation based on the same excitation frequency of 15.8Hz. (d) The corresponding time response of $\theta_{v}$ by numerically solving the equation of motion \ref{eq15}.  (e, f) The base acceleration and time response ($\theta_{v}$) from a similar numerical simulation using the analytically predicted resonance frequency (17Hz).}
    \label{fig:W2S}
\end{figure}

To experimentally validate this dynamic folding. We mount the origami on the shaker and manually set it at the weak stable state initially (Figure \ref{fig:test}(b)). We excite the shaker with a constant frequency of $\Omega=15.8$ Hz, which is the experimentally measured intra-well resonance frequency at this stable state.  Then, we gradually increase the amplitude of base excitation until the water-bomb ``snap'' to the strong stable state.  Once the snap occurs, we stop increasing the excitation amplitude (Figure \ref{fig:W2S}(a)).  The water-bomb origami continues to oscillate around the strong state without switching back to its original configuration.  We also replicate the same scenario numerically (Figure \ref{fig:W2S}(c, d). In this simulation, the excitation frequency equals to the experimentally measured resonance frequency, and the excitation amplitude increases linearly over time until the snap-through occurs.  It is worth noting that the numerical model predicts a higher base excitation amplitude required for switching.  This difference is due to the over-prediction of resonance frequency by the analytical model, as we discussed in the previous subsection.  In a different study shown in Figure \ref{fig:W2S}(e,f), we repeat the simulation exactly with the numerically predicted resonance frequency ($\Omega=17$ Hz), and observe a much smaller excitation amplitude requirement for switching.  Despite these quantitative differences, our model and experiment confirm the feasibility of dynamic folding from the weak stable state to the strong state solely by inducing an intra-well resonance with a small excitation amplitude. Moreover, we can reduce the required excitation magnitude by using this dynamic folding method.  That is, the required base displacement to achieve a dynamic folding from the weak to the strong stable state is $A=1.3$ mm, while the required base displacement is significantly higher if we fold the water-bomb quasi-statically ($A=6.5$ mm, Table \ref{tab:Comparison}).

\begin{table}
  \caption{Comparison between the required quasi-static displacement and dynamic excitation amplitude for the dynamic folding between two stable states. Here, the quasi-static displacement is based on the reaction force-displacement curve shown in Figure \ref{fig:FD}.\\[0.2in]}
    \label{tab:Comparison}
    \begin{tabular}{c c c c}
         \hline
         & \textbf{Quasi-static} & \textbf{Dynamic} \\
         \hline
         \textbf{Weak to Strong} & 6.5mm  & 1.3mm \\ 
         \textbf{Strong to Weak} & 51.1mm & 24.2mm \\
         \hline
    \end{tabular}
\end{table}

\subsection{Dynamic folding from the strong stable state}\label{Section 3_sub 3}

\begin{figure}[t!]
    \includegraphics[width=1.0\textwidth]{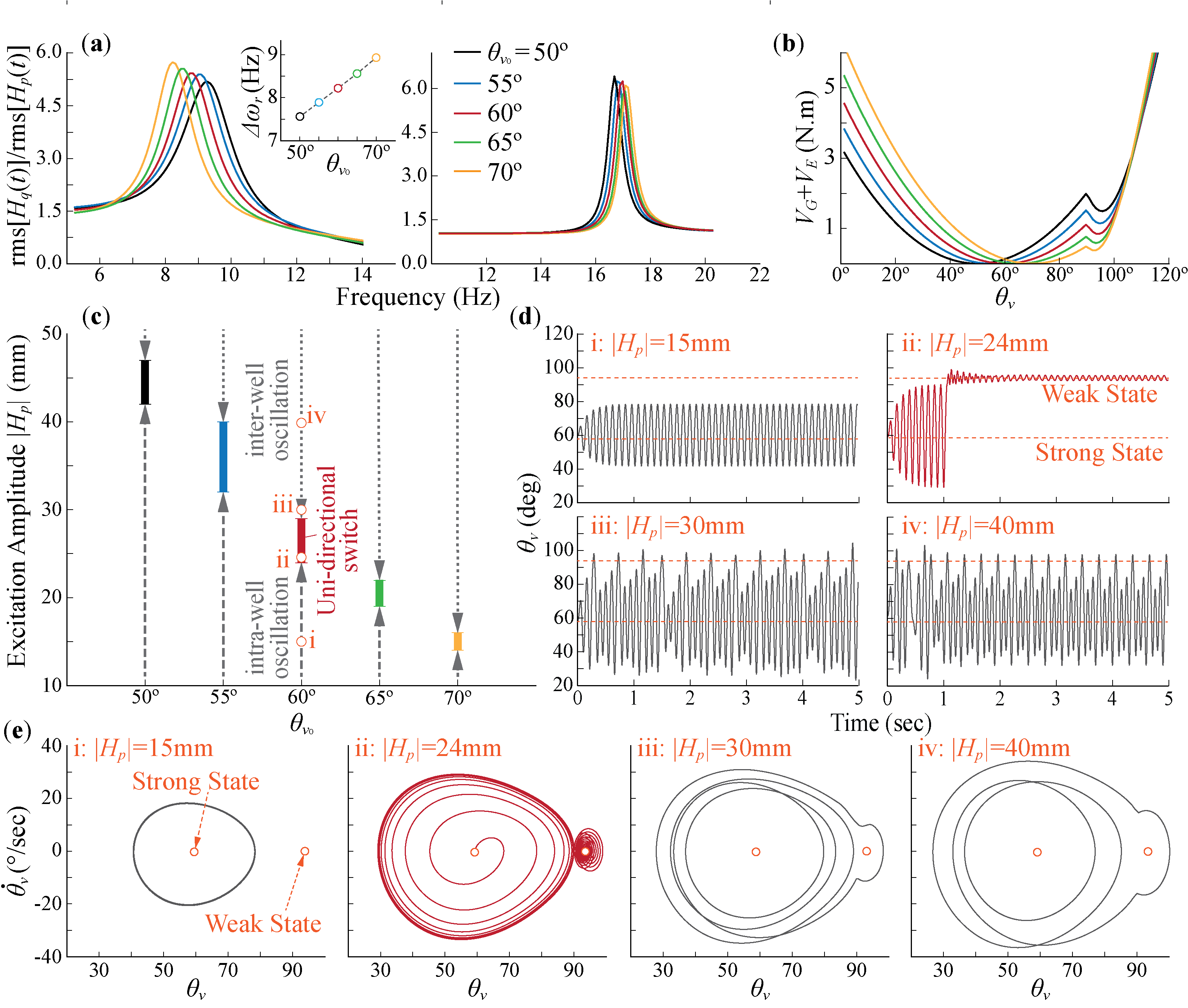}
    \caption{Dynamic folding behaviors of the water-bomb base origami from its strong stable state: (a) The numerically predicted frequency responses of water-bomb base origami structures with different stress-free folding angle $\theta_{v_{0}}$. The small insert figure shows the differences in resonance frequency between the two stable states. (b) The corresponding potential energy landscape.  (b) The correlation between stress-free folding angle, excitation amplitude, and the overall response.  The desired rapid folding (aka. uni-directional switch) is highlighted.  (e) The time responses of four representative cases based on different excitation amplitudes.  (e) The corresponding Poincare's map.  Note that, except for the case (ii), only steady-state responses are shown in these maps.}
    \label{fig:S2W}
\end{figure}

If the water-bomb base origami structure initially settles at the strong stable state, it has to overcome a significantly higher potential energy barrier $\Delta V_{G_1}$ to fold to the weak state. Although the intra-well resonance can help the origami to overcome this significant energy barrier, a large amount of energy in the system may lead to an \emph{inter-}well oscillation between the two stable states, which is not desirable for the dynamic folding purpose. 
To demonstrate this complex nonlinear dynamics, we conduct a parametric study to examine the relationships among the dynamic folding behaviors from the strong stable state, potential energy barriers, difference in resonance frequencies, and excitation amplitudes.  Figure \ref{fig:S2W}(a) shows the numerically predicted frequency responses of water-bomb origami with different stress-free folding angles ($\theta_{v_{0}}$) around their two stable states, while Figure \ref{fig:S2W}(b) shows the corresponding elastic potential energy landscape.  

We then excite each water-bomb origami with the resonance frequency of its strong stable state for a range of excitation amplitudes, all from zero initial conditions.  Figure \ref{fig:S2W}(c) summarizes their overall dynamic behaviors.  For every water-bomb design, there exists a small span of excitation amplitude that can generate the desired uni-directional switch (aka. rapidly folding from the strong state to the weak state without switching back).  For example, the case (ii) in Figure \ref{fig:S2W}(c)---with $\theta_{v_{0}}=60^\circ$ and $A=24$ mm---exhibits such a dynamic response.  Its time response and the corresponding Poincare's map are shown in figure \ref{fig:S2W}(d) and (e), respectively. One can observe that the oscillations of this water-bomb origami start from near the strong state, but eventually switch to and remain at the weak stable state.  

Any excitation below this span of uni-directional switch is not sufficient to overcome the potential energy barrier, leading to intra-well oscillations only (e.g., the case (i) in \ref{fig:S2W}(c-e) with $A=24$ mm).  On the other hand, any excitation above this span would generate an inter-well oscillation.  Case (iii) and (iv) \ref{fig:S2W}(c-d) are two examples of this inter-well oscillations.  Moreover, one can observe that, although both these two cases show inter-well oscillation, their state-state responses still show marked differences.  For example, a period in the steady-state response of case (iii) ($A=30$ mm) consists of three oscillations around the strong state before one inter-well oscillation, while the responses of the case (iv) ($A=40$ mm) only involve two oscillations around the strong state before an inter-well oscillation (Figure \ref{fig:S2W}(d, e)). 

Moreover, there is a clear trade-off between the potential energy barriers and natural frequency differences. As the stress-free folding angle $\theta_{v_{0}}$ of the water-bomb increases from $50^\circ$ to $70^\circ$, the difference in resonance frequencies also increases between the two stable states, however, the energy barrier $\Delta V_{G_2}$ decreases.  Therefore, as $\theta_{v_{0}}$ increases, the excitation magnitude corresponding to these spans of uni-directional switch decreases, and the width of these spans increases and then decreases.  Overall, we observe that a water-bomb origami with $\theta_{v_{0}}=55^\circ$ has the most balanced design and the widest excitation span to achieve a uni-directional switch.

Overall, our numerical simulations show that solely using the intra-well resonance to achieve the dynamic folding from the strong stable state to the weak state is possible but quite challenging.  That is, the excitation magnitude spans of the uni-directional switch are always narrow ($<10mm$) even with the more optimized origami designs.  Moreover, the nonlinear dynamics of the water-bomb base origami are quite sensitive to other uncertainties like initial conditions, fabrication errors, and excessive damping. For example, the actual differences in resonance frequencies are actually less than the prediction shown in Figure \ref{fig:resonance}.  As a result, we could not achieve a consistent and repeatable fold from the strong stable state to the weak one in the experimental efforts, despite the relatively small differences between the frequency response obtained from experiment and the prediction from numerical simulation. This challenge necessitates an active control strategy, as we detail in the next section.

\section{Active control strategy for robust folding}\label{Section 4}

\begin{figure}
    \includegraphics[width=1.0\textwidth]{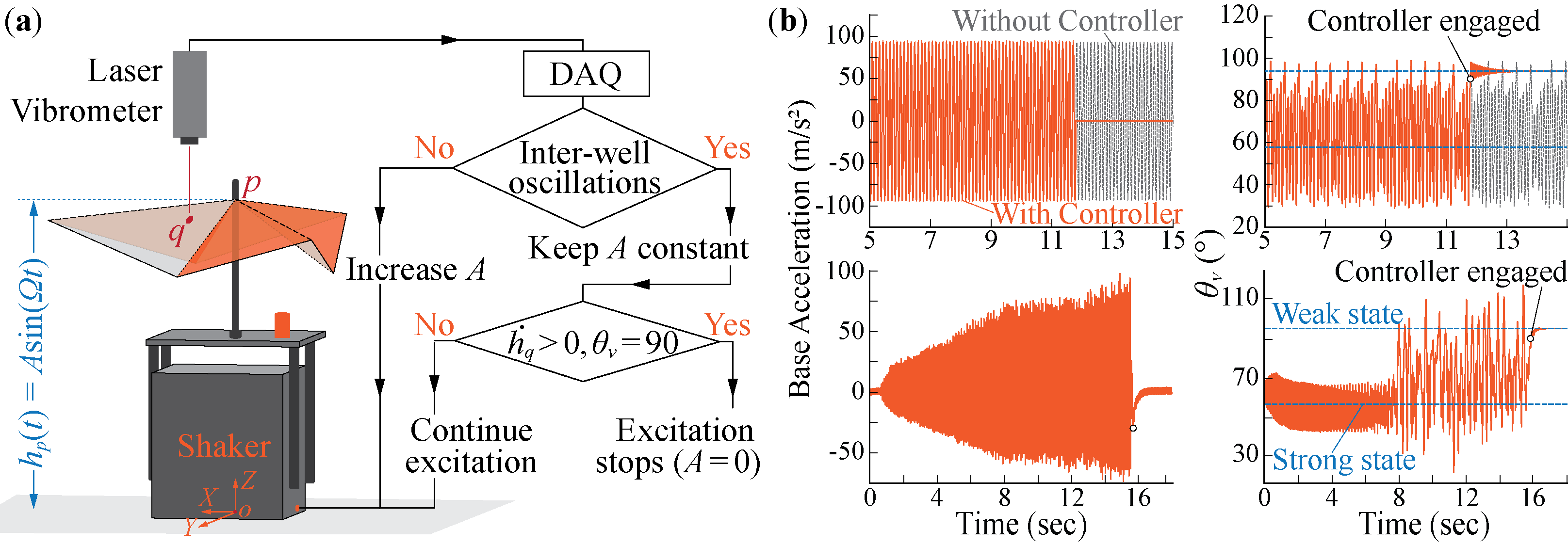}
    \caption{The control strategy to achieve a robust and dynamic folding from the strong stable state to the weak state. (a) The flow chart showing the concept and implementation of the controller (b) The controlled base acceleration and water-bomb origami folding angle based on the numerical simulation (top) and experimental validation (below).  It is clear that when the controller is engaged, inter-well oscillation is stopped quickly, and the water-bomb settles at the targeted weak stable state.}
    \label{fig:control}
\end{figure}

In this section, we propose a feedback control strategy that enables us to achieve a robust dynamic folding from the strong stable state of water-bomb origami to the weak state.  We show that this strategy is successful when pure dynamic excitation without control only generates inter-well oscillations between the two stable states. The idea of this feedback control strategy seems relatively straightforward.  Assuming the water-bomb origami is showing inter-well oscillations due to base excitation, we can cut off this excitation at the moment when the origami is folding toward the weak stable state (aka. $\dot{h}_q>0$) \emph{and} passing through the flat, unstable equilibrium (aka. $\theta_v=90^\circ$). In this way, the water-bomb origami should be able to overcome the energy barrier and switch to the weak stable state, but it would not be able to switch back to the strong state due to energy dissipation via damping.  Figure \ref{fig:control}(b) shows the numerically simulated folding with this controller.

We experimentally validate the effectiveness of this control strategy on the same water-bomb origami prototype. Figure \ref{fig:control}(a) shows the flow chart of this feedback loop based on the proposed control strategy. This feedback loop is encoded in a LabVIEW program that uses the laser vibrometer and accelerometer readings as the inputs. The labVIEW program filters out the acceleration data from the accelerometer using a bandpass filter and then integrates it twice to derive the displacement of the shaker's base. Then it calculates the relative displacement of the water-bomb base origami and the shaker by subtracting the derived displacements from the laser vibrometer readings. Finally, it calculates $\theta_v$ using this displacement data.  In this setup, we excite the water-bomb origami with the intra-well resonance frequency of $8.8$Hz and increase the excitation amplitude until an inter-well oscillation occurs. We then activate the controller, which can automatically detect the threshold of $\dot{h}_q>0$ and $\theta_v=0$ and cut of the excitation accordingly.  In our experiment, this controller can reliably and repeatedly fold the water-bomb origami from the strong stable state to the weak one (see supplemental video).  Therefore, despite its simplicity, the proposed controller provides an effective approach to complete the bi-directional dynamic folding of the water-bomb origami. Moreover, it is worth noting that the required base displacement to achieve dynamic folding from strong state to the weak state is $A=24.5$mm, which is much smaller than the excitation amplitude in a quasi-static folding ($A=51$mm, Table \ref{tab:Comparison}). It is also worth noting that this control algorithm is effective, but can certainly be modified further to increase its efficiency.

\section{Summary and Discussion}\label{Section 5}
In this study, we examine a dynamic and reversible origami folding method by exploiting the combination of resonance excitation, asymmetric multi-stability, and an active control strategy.   The underlying idea is that, by exciting a multi-stable origami at its resonance frequencies, one can induce rapid folding between its different stable equilibria without the need for responsive materials.  To this end, we use a bi-stable water-bomb base origami as the archetypal example and, for the first time, formulate a distributed mass-spring model to describe its nonlinear dynamics.  Via numerical simulations based on this new model and experimental testing using a proof-of-concept prototype, we characterize the difference in resonance frequencies between the two stable equilibria of the origami.  This difference stems from the inherent asymmetry of the water-bomb with respect to its unstable equilibrium at the unfolded flat shape.  We show that this asymmetry can enable dynamic folding with relatively low actuation requirements.   For example, if the water-bomb initially settles at its weak stable state, one can use a base excitation to induce the intra-well resonance.  As a result, the origami would fold and remain at the other stable state even if the excitation does not stop.  The origami dynamics near the strong state, on the other hand, is more complicated.  The asymmetric energy barrier makes the origami prone to show inter-well oscillation rather than a uni-direction switch.  There exist a complex trade-off between the desired uni-directional folding, potential energy barrier, the difference in resonance frequencies, and excitation amplitude. 

Therefore, we propose an active feedback control strategy to achieve robust and uni-directional folding from the strong stable state to the weak one.  This strategy cuts off the base excitation input when critical dynamic conditions occur.  Despite its simplicity, the control strategy is effective for controlling the dynamic folding.  We should emphasize that the proposed algorithm can be further modified to enhance performance.  For example, we can fully automate the task of detecting inter-well oscillations and sending control signals to cut off shaker input when necessary.  

It is worth noting that, although the results of this study are based on a water-bomb base origami, the physical insights into dynamic folding and the control strategy certainly apply to other origami or even other structures with asymmetric multi-stability.  It is also worth noting that, although this study assumes an external actuation, the underlying principles are still valid with embedded actuators inside the origami (e.g., along the creases).  Therefore, the results of this study can create a new approach for rapid and reversible (self-)folding, thus advancing the application of origami in shape morphing systems, adaptive structures, and reconfigurable robotics.

\begin{acknowledgments}
The authors acknowledge the support by the National Science Foundation (Award \# CMMI-1633952, 1751449 CAREER) and Clemson University (via startup funding and Deans's faculty fellow award).
\end{acknowledgments}

%\end{linenumbers}

\nocite{*}

\bibliography{resources}% Produces the bibliography via BibTeX.

\end{document}